\documentclass[10pt,twocolumn,jcp,aip,final,superscriptaddress,amsmath,amssymb,sort&compress,floatfix]{revtex4-1}

\usepackage{graphicx}

\usepackage{dsfont}

\begin{document}

\title{Freezing of parallel hard cubes with rounded edges}

\author{Matthieu Marechal}
\author{Urs Zimmermann}
\author{Hartmut L\"owen}
\affiliation{%
Institut f\"ur Theoretische Physik II: Weiche Materie,
Heinrich-Heine-Universit\"at
D\"usseldorf, Universit\"atsstra\ss e 1, D-40225 D\"usseldorf, Germany\\
}%

\date{\today}

\begin{abstract} 
The freezing transition in a classical 
three-dimensional system of  parallel hard cubes 
with rounded edges is studied by computer simulation and
fundamental-measure density functional theory. By switching the rounding parameter $s$
from zero to one, one can smoothly interpolate between
cubes with sharp edges and hard spheres. The 
equilibrium phase diagram of rounded parallel hard cubes is computed
as a function of their volume fraction and the rounding parameter $s$. The second 
order freezing transition known for oriented cubes at $s=0$
is found to be persistent up to $s = 0.65$. The fluid freezes
into a simple-cubic crystal which exhibits
a large vacancy concentration. Upon a further increase of $s$,
the continuous freezing is replaced by a first-order transition into 
either a sheared simple cubic lattice or a deformed face-centered cubic lattice with two possible unit cells: body-centered
orthorhombic or base-centered monoclinic.
In principle, a system of parallel cubes could be realized in experiments on colloids using advanced
synthesis techniques and a combination of external fields.\end{abstract}


\maketitle

\section{Introduction}

Most liquids freeze into a 
regular period crystalline lattice upon a sufficient temperature decrease or pressure increase.
Since this transition is associated 
with a  breaking of translational symmetry, it is typically discontinuous
(or first-order).\cite{Brazovskii1975}
This is in marked contrast to two spatial 
dimensions where no long-range translational order exists
\cite{Froehlich_Pfister} and the liquid-solid transition can be continuous, 
for example following the two-stage Kosterlitz-Thouless-Halperin-Nelson-Young 
scenario~\cite{[{For a review, see: }][{}]KTHNY}. One of the few (if not the only) exception to the common finding
of first-order freezing in three dimensions is a system of parallel hard cubes
where a disordered liquid freezes continuously into a simple cubic ({\it sc}) lattice at a volume fraction
of about 50 percent. 
This finding was first suggested for parallel hypercubes in more than three spatial dimensions
by Kirkpatrick \cite{Kirkpatrick} by using a second order virial expansion. Later,
Cuesta and coworkers
\cite{Cuesta_paraPRL1,Cuesta1997I,Cuesta_paraPRL2,martinez-raton1999} applied 
fundamental-measure density functional theory \cite{Rosenfeld1989FMTPRL,FMT2}
to parallel hard cubes in three dimensions showing that the freezing transition
is also continuous in three dimensions.
The continuous nature
 of the freezing transition was also confirmed later by computer simulations 
\cite{Jagla1998,groh2001cubes} where the associated criticality was found to be consistent with the 
Heisenberg universality class.\cite{groh2001cubes}

In this paper, we consider a more general shape of hard particles, 
namely cubes with rounded edges.
Our motivation to consider rounded cubes is threefold:
First, it is interesting how persistent
 the  second-order transition is with respect to a change of parameters. It is known
that orientable cubes (i.e. cubes with full orientational degrees of freedom) 
freeze via a first-order phase transition \cite{Jagla1998,Smallenburg_MM_cubes} such that 
additional rotational degrees lead back to the normal picture of freezing.
Therefore it is interesting to which extent the degree of rounding affects 
the order of the transition, in particular, since the extreme hard-sphere 
limit $d\to l$ exhibits the common first-order freezing scenario.

Second, fundamental-measure density functional theory (FMT) was developed \cite{Hansen-Goos2009edFMTPRL,Hansen-Goos2010edFMTlong}
and applied \cite{Haertel2010,Haertel2010PRE} further in recent years 
towards hard bodies of arbitrary shape. As a version of FMT already exists for the limiting cases, parallel hard cubes and hard spheres,
rounded cubes constitute an excellent 
model system to test the performance of fundamental-measure density functional theory.

Last but not least, it is now possible to fabricate micron-sized colloidal particles with almost arbitrary 
shapes \cite{Sun,Manoharan,glotzer2007anisotropy,Sacanna2,[{For a review and classification of the different shapes, see: }]Witt}. 
In a recent pioneering work of Rossi \emph{et al},\cite{Rossi2011rcubes}
well-defined colloidal cubes were prepared and studied in real-space by confocal microscopy.
These suspensions nicely realize the hard cube model of classical statistical mechanics.  
In the experimental samples, however, the cubes
 typically possess rounded edges, therefore our model shape model is closer 
to these colloids than the hard cube. In the experiments,
the colloidal cubes are not oriented in a parallel fashion. Furthermore, non-adsorbing polymers
were added to speed up the crystallization process. Therefore, a \emph{first}-order freezing transition
was found in this suspension.
Colloidal cubes can in principle be oriented by external aligning 
fields \cite{Hernandez-Navarro2011,Kim_align_mag} for instance by introducing an inner core with two distinct non-parallel dipole moments, each of which couples 
to a separate external field. 
By simultaneously applying two non-parallel external fields, which could be external electric or magnetic fields or a light field,
the orientation of the particle described by its two independent axes can be fixed.
We should note here,
that the phase behavior of a system of parallel monodisperse particles with only hard-core interactions at constant packing fraction is invariant under scaling of a dimension 
of the particle by a constant factor. Therefore, it is not inconceivable that our model of parallel rounded cubes would be realized in experiments, possibly
as stretched rounded cubes.
Furthermore, our work on parallel, rounded, hard cubes provides a good starting point for further studies on colloidal 
rounded cubes that are not aligned by external fields.

The model for a rounded cube that we use is a spherocube,
which can be obtained by rounding a cube with edge length $l$. 
The rounding is done by replacing all edges by quartered cylinders
 of diameter $d$ and the corners by a spherical octant such that the 
curvature is continuous on the cube's surface, see Fig.~\ref{fig:shape}.
If the diameter $d$ is zero, 
the traditional model of parallel hard cubes is recovered while for $0<d<l$
we are dealing with truly rounded cubes. Finally, in the extreme limit $d\to l$ we 
obtain the hard sphere model where freezing is known to be a first-order
transition into a face-centered-cubic ({\it fcc}) lattice.\cite{Lowen_PR} 
By splitting the particle surface
 into planar, cylindrical and spherical parts,  we propose 
a continuous interpolation between a cube and a sphere which is 
similar in spirit but different in practice to the superball 
interpolation used recently by Batten et al.\cite{Batten} 
To abbreviate the notation we define a rounding parameter $s=d/l$,
similar to Batten et al.'s $1/q$ for the superballs, in the sense 
that both $s=1$ and $1/q=1$ denote a sphere, while $s=0$ and $1/q=0$ denotes a cube.
The overlaps between two superballs can only be detected using an involved numerical algorithm,\cite{Batten} which leads to numerical
difficulties 
as the superball's shape approaches that of a cube.\cite{Batten} In contrast, the overlap algorithm for parallel spherocubes 
can be given in a closed and very simple form, as we will show in Appendix~\ref{sec:overlap_and_collision}.
Furthermore, the spherocube is a very convenient model particle for FMT, since the curvatures  that feature in the theory are constant on the spherical, cylindrical and planar sections
of the particle's surface. 
Therefore, we have chosen to use the spherocube as our model rounded cube instead of the superball.

We explore the rounded parallel cube model by Monte Carlo (MC) \cite{FrenkelSmit} and event-driven Molecular Dynamics (EDMD) \cite{Rapaport_bin_tree} computer simulations
and by fundamental measure density functional theory of freezing \cite{Rosenfeld1989FMTPRL,FMT2}
adjusted conveniently to the rounded shape.
As a simulation result, we calculate the 
equilibrium phase diagram of  rounded parallel hard cubes
as a function of packing fraction and the degree of rounding embodied in the 
ratio $s=d/l$. The second 
order freezing transition known for oriented cubes at $s=0$
is found to be very persistent occurring  up to high rounding degrees of about 
$s= 0.65$. This gives evidence that the second-order freezing transition 
can be seen in experiments on rounded oriented particles. The fluid freezes
into a simple-cubic crystal which is accompanied by
a very large vacancy concentration in the emerging solid.
At further increasing ratios $d/l$,
freezing becomes  a first-order transition into 
a sheared {\it sc} lattice and a deformed {\it fcc} lattice, where the
latter can have both orthorhombic (ortho) and monoclinic (clino) unit cells. 
Our simulation data for the continuous transition line and for the 
associated vacancy concentration are found to be
 in qualitative and semi-quantitative agreement with 
fundamental-measure density functional theory.
The three novel crystals (sheared {\it sc} and the ortho and clino variants of deformed {\it fcc})
can also be confirmed experimentally and could be useful for designing new 
materials with novel optical and rheological properties.

The paper is organized as follows: in section \ref{sec:model}, we introduce the rounded 
cube model in detail. We describe the simulation technique in 
section \ref{sec:sims} while providing the background of fundamental-measure 
density functional theory in section \ref{sec:theory}. Results are presented in 
section \ref{sec:results} and we conclude in section \ref{sec:conclusions}.

\section{The model of rounded parallel hard cubes} \label{sec:model}

\begin{figure}
\vspace*{1em}
\includegraphics[width=0.4\textwidth]{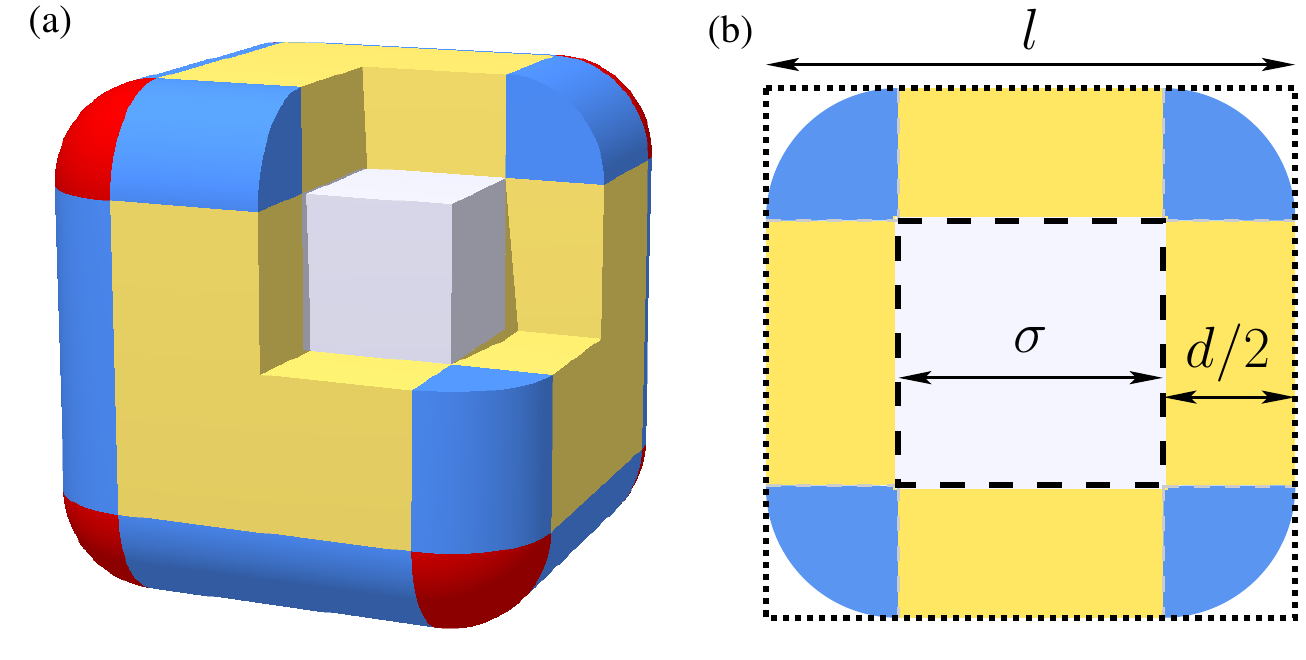}
\caption{(a) A spherocube or rounded cube consists of a cube (lightest/gray) surrounded by 6 square prisms (darker/yellow), 12 cylinder sections (still
darker/light blue) and 8 spherical sections (darkest/red).
   Some sections of the outer objects have been removed to show the gray cube. (b) Cross section of the spherocube showing
the edge length $\sigma$, minimum radius of curvature $d/2$, and the total width $l$.
 \label{fig:shape}}
\end{figure}

Our model rounded cube, the spherocube, is a special case of the sphero-cuboid introduced by Mulder in the context of second order virial theory.\cite{Mulder_ohsc}  
A spherocube can be obtained 
by coating a cube with edge length $\sigma$ with a layer of thickness $d/2$, as shown in Fig.~\ref{fig:shape}.
Alternatively, it can be constructed by rounding a larger cube with edge length $l=\sigma+d$ (dotted rectangle in Fig.~\ref{fig:shape}b),
such that its edges obtain a nonzero radius of curvature $d/2$. We will use $s\equiv d/l$ as the shape parameter for the spherocubes.
The volume of a rounded cube or spherocube is given by
\begin{equation}
v_\text{rc}=\frac{\pi}{6}d^3+\frac{3\pi}{4}d^2\sigma+\sigma^3+3 d\sigma^2.
\end{equation}
In Appendix~\ref{sec:overlap_and_collision}, we present the overlap algorithm for parallel spherocubes, which is surprisingly simple, especially compared to the overlap algorithm
for superballs.\cite{Batten}

The thermodynamic state of the system is sometimes specified using the pressure $P$,
but mostly using the volume fraction or packing fraction $\eta\equiv v_\text{rc} \rho=v_\text{rc} N/V$, where $\rho$ is the density, $N$ the number of particles
and $V$ the volume of the system. The temperature $T$
only serves to define the energy unit $k_B T$, where $k_B$ is Boltzmann's constant.

\section{Computer simulations}  \label{sec:sims}

In this section, which consists of four parts, the simulations that were performed in this work are described.
First, we determined candidate crystal structures using a recent, but well-tested
simulation technique,\cite{PhysRevLettSSS} as summarized in the
first part of this section.
After that we describe the Monte Carlo (MC) and event-driven Molecular dynamics (EDMD) techniques.
The structural and thermodynamic properties we measure are listed in the third part
and, finally, we describe the methods used to determine the phase behavior in Sec.~\ref{subsec:phbeh}.

\subsection{Candidate crystal structures} \label{subsec:uc_sims}

We find candidate crystal structures by simulating a single unit cell with fully variable box lengths and angles~\cite{Parrinello,Najafabadi1983} in the $NPT$ ensemble, that is the number of
particles $N$, the pressure $P$ and the temperature $T$ are held fixed.
Using periodic boundary
conditions, this unit cell is replicated indefinitely to roughly approximate an infinite crystal.
The final configurations of a number of compression series form the unit cells of the potentially stable crystal structures.
This computationally inexpensive method has been shown to find all stable crystal phases when applied to a system where the phase behavior
was already known~\cite{PhysRevLettSSS} and since then has been employed to find candidate structures for a number of novel systems
~\cite{MM_Kortschot2010bowls,MM_Dijkstra2010full_bowls,MM_Cuetos2011platelets}
and also to find close packed structures.\cite{De_Graaf_packing}

The variant of the method we use is the following: We run a large number of fast compression runs, see Ref.~\onlinecite{MM_Dijkstra2010full_bowls} for details. 
At the lower pressures, the system samples many meta-stable states. As the pressure is quickly increased, the system gets stuck in one
of these states. Finally, a nearly perfect crystal is found at very high pressure. To distinguish between different crystals we use
the box shape parameter introduced by De Graaf \emph{et al},\cite{De_Graaf_packing}, which is the average length of the box edges times the average area 
of its faces divided by its volume. The states are divided in clusters, such that the box shape parameter of each state in a certain cluster
deviates less than $10^{-4}$ from the box shape parameter of at least one other state in the cluster. The state which used to represent
the cluster is the state with the highest density, as this will often be the most ordered one. 
In the remainder of this work we will refer to this method for determining candidate crystal structures as ``unit cell simulations''.

The small system size allows large fluctuations in parameters such as the density, which ensure that all possible
states are visited.
However, the small system size would lead to huge finite size effects, if the results from these simulations would be directly
used to determine the region of stability and other thermodynamic properties of the crystals that were found.
Rather, this method is intended to be used in concert with conventional simulations (see below),
which take crystals formed by replicating the unit cells obtained from this method as initial configurations.

\subsection{Simulation techniques}\label{sim:tech}

We implemented EDMD simulations for hard rounded cubes, which allows us 
to measure the pressure very efficiently and to quickly equilibrate the system.
Molecular dynamics for hard particles is implemented by solving the 
equation of motion exactly. As such, hard particles perform free motion
interrupted by instantaneous collisions in EDMD.
Event driven MD simulations are especially fast when collisions can be predicted analytically, such as for hard spheres, and 
also for the rounded cubes studied in this work,
as described in Appendix~\ref{sec:overlap_and_collision}. 

Although event driven MD simulations are very fast, we found it to be more convenient to use Monte Carlo (MC) simulations in the following situations:
Many non-cubic crystals show a deformation of the unit cell upon a change in density or pressure.
Moves that change the shape of the box~\cite{Parrinello} can be easily added to Monte Carlo simulations~\cite{Najafabadi1983} to account\
for these deformations.
Furthermore, external potentials,
such as the ones required for the free energy calculations described further on, can easily be accounted for in
Monte Carlo simulations, while they would make the free motion in between collisions too complicated to predict the collisions analytically
in EDMD simulations.
Finally, we want to allow the vacancy concentration to adjust to changes in the density or pressure. The simplest way
to allow the vacancy concentration to change is to allow the box to change its shape in a Monte Carlo simulation. A simulation box which
starts with $M_0\equiv N^{(0)}_x\times N^{(0)}_y\times N^{(0)}_z$ unit cells with one particle in each unit cell can transform 
into $M \equiv N_x\times N_y\times N_z$ cells for any integers $N_x$, $N_y$ and $N_z$, such that $M>M_0$. The resulting vacancy concentration 
is $\nu_\text{vac}\equiv 1-M_0/M$.
In practice, we use a $N^{(0)}_z$ which is 50 or 100, such that the vacancy concentration is at lowest
$1-100/101\simeq 0.01$ or $1-50/51\simeq 0.02$. Note, that the minimal vacancy concentration is an order of magnitude larger than the vacancy concentrations in common crystals
(for instance
the vacancy concentration is of order $10^{-4}$ for hard spheres~\cite{Bennett_vacancies,Oettel_Schilling_MC_FMT_HS}).
However, the vacancy concentrations in simple cubic crystals of rounded cubes are orders of magnitude larger than those of hard spheres,
as we will see below, which allows us to use this simple technique to measure $\nu_\text{vac}$.
However, to keep the run time of the simulation limited we have to use considerably smaller
system sizes in the other directions: $N_x,N_y$ are either 10 or 15. 
In the other Monte Carlo simulations, we have used a system of approximately 1000 particles unless mentioned otherwise.

\subsection{Thermodynamic and structural properties} \label{subsec:quants}

The order parameter $m$ which measures the degree of crystallinity was introduced by Groh and Mulder.\cite{groh2001cubes}
It is defined using the maximum of the Fourier transformed density profile for $m_\nu$, where $\nu=x,y,z$ denotes a direction along one of the Cartesian axis
and where the density profile is averaged over the two other directions before performing the Fourier transform:
\begin{equation}
m_\nu=\max_{k} \hat{\rho}_\nu(k).\label{eqn:m_nu}
\end{equation}
The order parameter $m$ is defined by $m=( \lvert m_x\rvert + \lvert m_y \rvert + \lvert m_z \rvert)/3$.
Because the vacancy concentration and therefore the number of unit cells in each direction can change in the variable box length NPT
MC simulations (see Sec.~\ref{sim:tech}), we do not know
the reciprocal lattice vector $k_\nu \hat{x}_\nu$, in the $\nu$-direction $\hat{x}_\nu$ before hand (However, we do know 
its direction, because the particles do not rotate). For this reason, we maximize with respect to $k$ in Eq.~\ref{eqn:m_nu}.
We can use the resulting length of the wave vector $k_\nu$ to determine the number of unit cells in the $\nu$ direction:
$N_\nu= k_\nu L_\nu/2\pi$, where $L_\nu$ is the length of the simulation box in the $\nu$ direction. Obtaining
$N_\nu$ in all three directions in this way, we can calculate the vacancy concentration $\nu_\text{vac}$ as $\nu_\text{vac}=1-N/(N_x N_y N_z)$. 

We measure the equation of state using one of two methods depending on the phase of interest.
The pressure of the fluid phase as a function of density is measured in $NVT$ EDMD simulations, 
while the density as a function of the pressure of each of the crystal phases is measured 
using $NPT$ MC simulations.

We also calculated the mean squared deviation from the lattice position (MSD) to compare to the FMT data. 
For system with vacancies, the obvious definition of this quantity gives infinity because the particles
can easily diffuse away from their lattice position by hopping to a neighboring, empty lattice site. 
Instead, we measure the MSD from the \emph{nearest} lattice site. However, we need to know the positions of the
perfect lattice sites; specifically, the shift $r_{\nu,0}$ of the lattice compared to the lattice which 
has one of its lattice sites in the origin. We use EDMD simulations with zero total momentum.
This means that, when a particle hops a lattice constant $a_0$
to, say, the left compared to the lattice, the lattice shifts $a_0/N$ to the right because the center of mass is fixed. 
As such the shift $\mathbf{r}_0$ drifts with time, and needs to be obtained in the simulation before the MSD can be measured.
A reliable way of determining the shift $\mathbf{r}_0$ is to use the phase of $\hat{m}_\nu(k_\nu)$, which should be equal
to $\exp(\mathrm{i} k_\nu r_{\nu,0})$. 

\subsection{Phase behavior} \label{subsec:phbeh}

We know from earlier work for perfect cubes~\cite{groh2001cubes} and spheres~\cite{Hoover_Ree} that the phase diagram contains both first and second order melting transitions.
For the first order transitions we use the highly accurate free energy methods that were developed by Frenkel and co-workers.\cite{frenkel1984einstein} 
We summarize these methods in Sec.~\ref{sec:free_E}. The second order phase transitions were located using finite size scaling~\cite{Binder1997appMC,groh2001cubes}
as described in the section after that.

\subsubsection{Free energy calculations} \label{sec:free_E}

The Helmholtz free energy of the fluid is obtained by integrating the equation of state from the ideal gas limit:
\begin{equation}
f^*_\text{fluid}(\rho)=\log(\rho l^3)+\int_0^\rho \mathrm{d} \rho (P/\rho-1) /\rho \label{eqn:f_fluid}
\end{equation}
where, here and in Appendix~\ref{sec:FrLadd_vac}, the free energy is made dimensionless by $f^*\equiv F/(N k_B T) - \log(\Lambda^3/l^3)$, $\Lambda$ is the (irrelevant) thermal
wavelength $\Lambda=h/\sqrt{2\pi m_\text{rc} k_B T}$, $m_\text{rc}$ is the mass of a particle and $h$ is Planck's constant.

\begin{figure}
\includegraphics[width=0.4\textwidth]{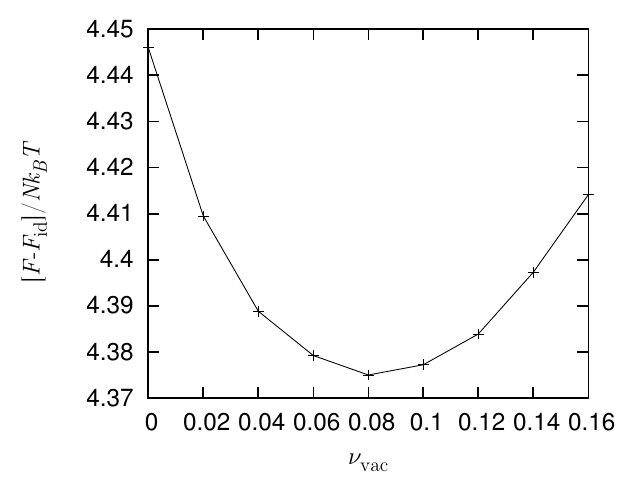}
\caption{
The free energy of a simple cubic crystal of spherocubes with $s=0.6$ as a function of its vacancy fraction $\nu_\text{vac}$ at 
$\eta=0.53$ for $N\simeq 1000$ particles. 
\label{fig:f_of_vac}}
\end{figure}

The free energies of the crystal phases are measured using the Frenkel-Ladd method~\cite{frenkel1984einstein,FrenkelSmit} in which the free energy difference between a
crystal and the non-interacting Einstein crystal is calculated by thermodynamic integration. 
We have made some modifications to the method when applying it to the simple cubic crystal phase to allow for a nonzero vacancy concentration.
As these modifications are similar to the one applied for rotating cubes in Ref.~\onlinecite{Smallenburg_MM_cubes}, we leave the 
details for Appendix~\ref{sec:FrLadd_vac}.
As an example, the free energy, resulting from the thermodynamic integration technique, is shown as a function of the 
vacancy concentration in Fig.~\ref{fig:f_of_vac}. Clearly, a finite (and quite large) vacancy concentration is found, around
8\%. This is surprising, because the packing fraction for this free  energy, $\eta=0.53$, is rather high compared to the critical density $\eta_c\simeq 0.47$,
as determined using the methods described below. We calculated two more free energy curves as a function of vacancy concentration and the 
resulting vacancy concentrations are shown together which the results from the variable box length $NVT$ simulations and the FMT in Sec.~\ref{sec:results}.
Minimizing the free energy with respect to vacancy concentration at every density is somewhat cumbersome, so we have used the variable box length simulations
to determine the vacancy fraction in most of this work as described in Sec.~\ref{subsec:quants}. 
Once a reference free energy $f^*(\rho_0)$ is known at a certain reference density $\rho_0$ for each crystal, we integrate over the
equation of state $P/\rho^2$ to obtain the free energy at all densities, similar to Eq.~(\ref{eqn:f_fluid}). 

When the free energies of all relevant phases at a certain aspect ratio $s$ are known,  the coexistence densities for a given pair of phases $1$ and $2$ can be found by solving 
$P_1(\rho_1)=P_2(\rho_2)$ and $\mu_1(\rho_1)=\mu_2(\rho_2)$, where the pressure $P_i$ of phase $i$ is obtained from a fit of the equation of state
and the chemical potential $\mu_i=F_i/N_i+P_i/\rho_i$. Solving these equations for all possible pairs of phases and finding at each density
the phase or phase coexistence, which has the lowest free energy, the phase diagram can be drawn.

\subsubsection{Finite size scaling}

For a second order phase transition, the above method can not be used due to very large finite size effects near the transition.
Near the phase transition, we can use finite size scaling~\cite{Binder1997appMC} to find the properties of the infinite system. 
Only the behavior of the order parameter with pressure and system size 
is required to find the location of the transition; no thermodynamic integration is required in this case. 
We use the scaling of $\langle m\rangle = N^{-\nu_1} \tilde{m}(\lvert P_c/P-1\rvert N^{\nu_2})$ and 
the Binder cumulant~\cite{Binder1997appMC}
$U_N\equiv 1-\langle m^4\rangle / \langle m^2 \rangle^2 =\tilde{U}(\lvert P_c/P-1\rvert N^{\nu_2})$, where $P_c$ is the critical pressure,
 $\nu_1$ and $\nu_2$ 
are finite size scaling exponents and $\tilde{m}$ and $\tilde{U}$ are scaling functions. The exponents $\nu_1$ and $\nu_2$ fall into certain 
universality classes (In terms of the critical exponents $\beta$ and $\nu$ often used in the literature, $\nu_1=-\beta/3\nu$ and $\nu_2=1/3\nu$).
Groh and Mulder~\cite{groh2001cubes} determined the universality class for the melting transition of hard cubes without vacancies
to be that of the three dimensional classical Heisenberg model, which has $\nu_1\simeq 0.173$ and $\nu_2\simeq 0.472$.\cite{Chen_exponents} We have not found any evidence
that these exponents are changed when vacancies are included and, therefore, use these values for the exponents also here.

We used the system sizes $N=10^3$, $15^3$ and $20^3$, which are large enough that we can neglect corrections to finite size scaling~\cite{groh2001cubes}
(our large system size is considerably larger than that of Ref.~\onlinecite{groh2001cubes}). 

\begin{figure}
\includegraphics[width=0.495\textwidth]{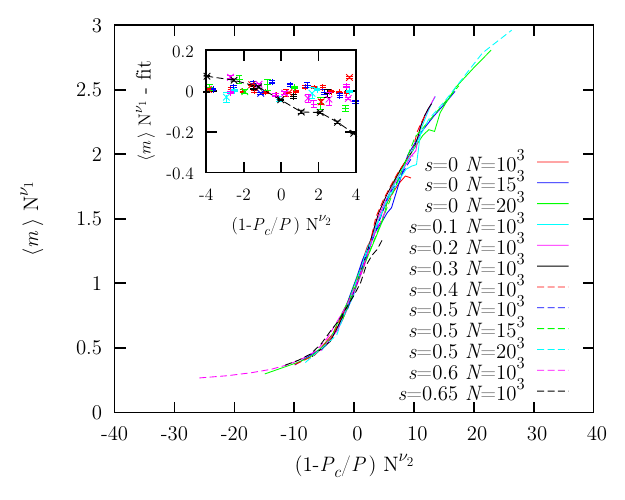}

\caption{
The scaled positional order parameter, $\langle m \rangle N^{\nu_1}$, of systems of $N$ hard rounded parallel cubes with
$s=0$, 0.1 0.2 0.3 0.4 0.5 0.6 and 0.65 as a function of
$(1-P/P_c) N^{\nu_2}$, where $P_c$ is the critical pressure
(that is a function of $s$) and $\nu_1$ and $\nu_2$ are scaling exponents.
The system size is $N=10^3$ for the results for $s=0$ and $s=0.5$, for which $N=10^3$, $15^3$ and $20^3$.
The inset shows $\langle m \rangle N^{\nu_1}$  near the
critical pressure with a fit to the results for $s=0$ and $N=10^3$ subtracted for clarity. The 
pluses and crosses correspond to the solid and dashed lines, respectively, of the same color in the main plot.
\label{fig:collapse}}

\end{figure}

A method to determine the critical pressure that does not use
the values of the scaling exponents consists of plotting $U_N$, which does not depend on system size for $P=P_c$, for a number of system sizes.
The point where the three curves meet is the critical pressure $P_c$. 
We used this method for $s=0$ and $s=0.5$. When $N^{\nu_1} \langle m\rangle$ is plotted against 
$\lvert P_c/P-1\rvert N^{\nu_2}$ the curves fall on top of a single master curve, which confirms the finite size scaling 
ansatz that $\tilde{m}(\lvert P_c/P-1\rvert N^{\nu_2})\equiv\langle m\rangle N^{\nu_1}$ is a universal function
of the relative deviation from the critical pressure. We exploited the universalness of this function to determine 
$P_c$ for the values $s=0.1$, $0.2$, $0.3$, $0.4$, $0.6$ and $0.65$ by fitting $P_c$ such that the data for $\tilde{m}(\lvert P_c/P-1\rvert N^{\nu_2})$
fall on top of a single master curve for all values of $s$ and system sizes considered.
The collapse is shown in Fig.~\ref{fig:collapse}. Note, that the vacancy concentration
changes as a function of the pressure. Consequently, the number of unit cells in a certain direction changes discretely in these simulations,
which is the cause of the noise in Fig.~\ref{fig:collapse}. Reassuringly, the collapse of the positional order parameters is reasonably good, considering the
noise. The inset shows a zoom near the critical pressure and a fit to the data for $s=0$ and $N=10^3$ has been subtracted for clarity.
Most of the data in the inset is indeed scattered around zero for a range of pressures near $P=P_c$, where the exception
seems to be $s=0.65$ (black dashed line).
Apparently, corrections to scaling are more important for this value of $s$, which is close to the triple point where the simple cubic
phase is replaced by another crystal phase.
Therefore, the critical pressure $P_c$ for $s=0.65$ is less accurate
than $P_c$ for the other values for $s$.

\section{Fundamental-measure density functional theory}  \label{sec:theory}

In the framework of density functional theory \cite{Evans} the equilibrium grand canonical potential is obtained by minimizing the density functional
\begin{align*}
\Omega[\rho] = \mathcal{F}[\rho] + \int \!\!\mathrm{d}{\mathbf{r}} \rho (\mathbf{r})\left(V_{\mathrm{ext}}(\mathbf{r}) - \mu\right),
\end{align*}
with the intrinsic free energy functional $\mathcal{F}[\rho]$, the external potential $V_{\mathrm{ext}}(\mathbf{r})$, the density distribution
$\rho(\mathbf{r})$ and the chemical
potential $\mu$. We limit our considerations to a one-component system but the theory can easily be generalized to multicomponent systems. The functional $\mathcal{F}[\rho]$ naturally separates into two parts, $\mathcal{F}[\rho] = \mathcal{F}_\text{id}[\rho] + \mathcal{F}_{\mathrm{exc}}[\rho]$ with the ideal gas contribution
\begin{align*}
	\mathcal{F}_\text{id}[\rho] = k_\mathrm{B} T \int \mathrm{d}{\mathbf{r}}\rho(\mathbf{r})\left(\log\left(\rho(\mathbf{r})\Lambda^3\right)-1\right),
\end{align*}
where $\Lambda$ is the thermal wavelength as defined above.
The excess free energy $\mathcal{F}_{\mathrm{exc}}[\rho]$ which contains the information of particle interactions is not exactly known, such that
one has to rely on approximations. 

For hard sphere systems, Rosenfeld's fundamental measure theory (FMT) \cite{Rosenfeld1989FMTPRL} and refined versions of FMT~\cite{Roth_WBI,Hansen-Goos2006WBII} are currently
the most accurate density functional approaches~\cite{[{For a review, see }][]Roth2010revFMT,*Tarazona_Cuesta_Rev_DFT}. Rosenfeld \cite{Rosenfeld1995} also generalized the FMT to arbitrarily convex shaped hard interacting particles
using the Gauss--Bonnet theorem. 
His theory yields good results only for mildly elongated particles~\cite{[{For an application of edFMT to dumbbells, see: }][{}]MM_Goetzke_db_FMT}.
Recently, Hansen--Goos and Mecke improved these
considerations \cite{Hansen-Goos2009edFMTPRL,Hansen-Goos2010edFMTlong} within the so--called extended deconvolution fundamental measure theory (edFMT). First, we
briefly review the results of edFMT and subsequently apply it to spherocubes.

\subsection{General approach of edFMT}

In the low density limit, we can express the excess free energy functional as a second order virial expansion:
\begin{align}\label{eq:ldl}
\lim_{\rho \rightarrow 0} \mathcal{F}_{\mathrm{exc}} [\rho] = -\frac{k_\mathrm{B} T}{2} \iint \mathrm{d}{\mathbf{r}'} \mathrm{d}{\mathbf{r}}\rho(\mathbf{r}') \rho(\mathbf{r}) f(\mathbf{r} - \mathbf{r}').
\end{align}
Here $f(\mathbf{r})=\exp\big(-\varphi(\mathbf{r})/k_\mathrm{B} T)\big) - 1$ is the Mayer function with pair interaction potential $\varphi$ between two particles. In the context
of hard--body interactions the Mayer function simply reads $f(\mathbf{r}) = -1$ for overlapping particles and $f(\mathbf{r}) = 0$ otherwise. The Mayer function
can be deconvoluted into a sum of weight functions $w_{\alpha}$ that capture the geometrical features of a single convex particle, as shown in
\cite{Hansen-Goos2009edFMTPRL,Hansen-Goos2010edFMTlong}:
\begin{align}\label{eq:Mayer}
	-\frac{f(\mathbf{r})}{2} &= w_{0}\otimes w_{3}(\mathbf{r}) + w_{1}\otimes w_{2}(\mathbf{r}) - \mathbf{w}_{1}\otimes \mathbf{w}_{2}(\mathbf{r}) \\
	&\quad- \sum_{j=2}^{\infty}(-1)^j {W}_{1}^{[j]} \otimes {W}_{2}^{[j]}(\mathbf{r}).\notag
\end{align}
Here and in the remainder, we will denote scalar quantities by $x$, vector quantities
by $\mathbf{x}$ and tensors of rank $j\geq 2$ by ${X}^{[j]}$.
The entire set of related scalar, vectorial and tensorial quantities is referred to as $\{x_\alpha\}$. The operation $\otimes$ of two weight functions is defined by
\begin{align*}
w_{\alpha} \otimes w_{\gamma}(\mathbf{r}) = \int \!\!\mathrm{d}{\mathbf{r}'} w_{\alpha}(\mathbf{r'}) \star w_{\gamma}(\mathbf{r'} - \mathbf{r}).
\end{align*}
Here, we use the generalized scalar product $\star$, which is meant to be a multiplication for scalar quantities, scalar product for vector quantities and trace of the product of two matrices for tensors of second order. In general, for tensors of $j$th order we have:
\begin{align*}
{X}^{[j]} \star {Y}^{[j]} &= \sum_{i_1, \dots, i_j} \left( {X}^{[j]} \right)_{i_1 \dots i_j}  \!\!\! \cdot \left( {Y}^{[j]} \right)_{i_j \dots i_1}.
\end{align*}
The geometrical weight functions $\{w_{\alpha}\}$ 
are given by
\begin{align*}
w_{3}(\mathbf{r}) &= \Theta \left( |\mathbf{R}(\mathbf{\hat{r}})| - |\mathbf{r}| \right)\\
w_{2}(\mathbf{r}) &= \frac{\delta \left( |\mathbf{R}(\mathbf{\hat{r})}| - |\mathbf{r}|\right)}{\mathbf{\hat{n}}\cdot \hat{\mathbf{r}}} \\
w_{1}(\mathbf{r}) &= \frac{H(\mathbf{r})}{4\pi} w_{2}(\mathbf{r})\\
w_{0}(\mathbf{r}) &= \frac{K(\mathbf{r})}{4\pi} w_{2}(\mathbf{r})\\
\mathbf{w}_{2}(\mathbf{r}) &= \hat{\mathbf{n}} w_{2}(\mathbf{r})\\
\mathbf{w}_{1}(\mathbf{r}) &= \frac{H(\mathbf{r})}{4\pi} \mathbf{w}_{2}(\mathbf{r})\\
{W}_{1}^{[2]}(\mathbf{r}) &= \frac{\Delta\kappa(\mathbf{r})}{4\pi} \left( \mathbf{v}^{\mathrm{\MakeUppercase{\romannumeral 1}}} {\mathbf{v}^{\mathrm{\MakeUppercase{\romannumeral 1}}}}^{T} \!\! (\mathbf{r}) - \mathbf{v}^{\mathrm{\MakeUppercase{\romannumeral 2}}} {\mathbf{v}^{\mathrm{\MakeUppercase{\romannumeral 2}}}}^{T} \!\! (\mathbf{r})\right) w_{2}(\mathbf{r}) \\
{W}_{2}^{[2]}(\mathbf{r}) &= \hat{\mathbf{n}}(\mathbf{r}) \hat{\mathbf{n}}^T \!\! (\mathbf{r}) w_{2}(\mathbf{r}) \\
\left({W}_{1}^{[j]}\right)_{i_1\dots i_{j}} \!\!\!\!\!\!\! &= \frac{\Delta \kappa(\mathbf{r})}{4\pi}\left( v_{i_1}^{\mathrm{\MakeUppercase{\romannumeral 1}}}v_{i_2}^{\mathrm{\MakeUppercase{\romannumeral 1}}} - v_{i_1}^{\mathrm{\MakeUppercase{\romannumeral 2}}}v_{i_2}^{\mathrm{\MakeUppercase{\romannumeral 2}}} \right)\hat{n}_{i_3}\!\!\! \cdots \hat{n}_{i_{j}} w_{2}(\mathbf{r}) \\
\left({W}_{2}^{[j]}\right)_{i_1\dots i_{j}} \!\!\!\!\!\!\! &= \hat{n}_{i_1}\hat{n}_{i_2} \cdots \hat{n}_{i_{j}} w_{2}(\mathbf{r}).
\end{align*}
In this notation, $\mathbf{R}(\mathbf{\hat{r}})$ is the vector that points along the direction $\mathbf{\hat{r}}$ from a certain reference point inside a
particle to its surface.
 The principle curvatures $\kappa_{\mathrm{\MakeUppercase{\romannumeral 1}}}$ and $\kappa_{\mathrm{\MakeUppercase{\romannumeral 2}}}$ with corresponding principle directions $\mathbf{v}^{\mathrm{\MakeUppercase{\romannumeral 1}}}$
and $\mathbf{v}^{\mathrm{\MakeUppercase{\romannumeral 2}}}$ are defined at $\mathbf{R}(\mathbf{\hat{r}})$. We also define the mean curvature \mbox{$H =
(\kappa_{\mathrm{\MakeUppercase{\romannumeral 1}}} + \kappa_{\mathrm{\MakeUppercase{\romannumeral 2}}})/2$}, the Gaussian curvature \mbox{$K = \kappa_{\mathrm{\MakeUppercase{\romannumeral 1}}} \kappa_{\mathrm{\MakeUppercase{\romannumeral 2}}}$} and the deviatoric curvature \mbox{$\Delta\kappa = (\kappa_{\mathrm{\MakeUppercase{\romannumeral 1}}} - \kappa_{\mathrm{\MakeUppercase{\romannumeral 2}}})/2$}
for convenience. The normal vector is written as $\hat{\mathbf{n}}$. With $\mathbf{a}^T$ we denote the transpose of the vector or matrix. Consequently, $\mathbf{a} \mathbf{b}^T$ is the dyadic product of vectors $\mathbf{a}$ and $\mathbf{b}$.

Finally, we introduce weighted densities as a convolution of the density profile with the corresponding weight function:
\begin{align}\label{eq:ndef}
 n_{\alpha}(\mathbf{r}) = \int \mathrm{d}{\mathbf{r}'}\rho(\mathbf{r'}) w_{\alpha}(\mathbf{r-r'}).
\end{align}
In terms of weighted densities the low density limit \eqref{eq:ldl} becomes
\begin{align}\label{eq:ldl_n}
\lim_{\rho \rightarrow 0} \mathcal{F}_{\mathrm{exc}} [\rho] &= k_\mathrm{B} T \int \mathrm{d}{\mathbf{r}} \left[ n_{0}n_{3} + n_{1}n_{2} - \mathbf{n}_{1} \cdot \mathbf{n}_{2}\right.\notag\\
& \qquad - \left. \sum_{j=2}^{J} \zeta_j {N}_{1}^{[j]} \star {N}_{2}^{[j]}\right].
\end{align}
The tensorial weighted density ${N}_\alpha^{[j]}$ should not be confused with the number of particles $N$.
Here, we truncated the tensor expansion after the $J$th term and introduced the free parameters $\zeta_j$. 
In the limit $J\to\infty$, the exact low density limit is recovered provided $\zeta_j=(-1)^j$.
The tensorial terms account for the asphericity and vanish only for hard sphere particles. In their 
original work on edFMT,\cite{Hansen-Goos2009edFMTPRL} Hansen--Goos and Mecke truncated the infinite sum at $J=2$. Subsequently, they showed,
for an anisotropic fluid of spherocylinders, that 
renormalization by $\zeta_2\neq 1$ better corrects for the influence of the truncation than including higher order tensorial terms without 
renormalizing (\emph{i.e.} with $\zeta_2=1$).\cite{Hansen-Goos2010edFMTlong}

In edFMT the following ansatz is made for the excess free energy
\begin{align*}
\mathcal{F}_{\mathrm{exc}}[\rho] = k_\mathrm{B} T\int \mathrm{d}{\mathbf{r}} \Phi[\{n_\alpha(\mathbf{r})\}],
\end{align*}
where the free energy density $\Phi$ solely depends on the set of weighted densities $\{n_\alpha\}$. Clearly, as $\rho \rightarrow 0$ we should recover \eqref{eq:ldl_n}.
As a result of scaled particle theory and dimensional analysis, the free energy density for truncated tensor terms reads
\begin{align}\label{eq:free_energy_density}
\Phi &= -n_0\log (1-n_3) + \frac{\phi_1(\{n_\alpha\})}{1-n_3} + \frac{\phi_2(\{n_2\})}{(1-n_3)^{2}},
\end{align}
with
\begin{align}
\phi_1(\{n_\alpha\}) &= n_1n_2 - \mathbf{n}_1 \cdot \mathbf{n}_2 - \sum_{j=2}^{J}\zeta_j {N}_{1}^{[j]} \star {N}_{2}^{[j]}\label{eq:phi1}\\
\phi_2(\{n_2\}) &=	c_0\left( \mathbf{n}_2^T {N}_{2}^{[2]} \mathbf{n}_2 - n_2\mathbf{n}_2\cdot\mathbf{n}_2 \right.\nonumber\\
&\qquad \left. + n_2 \operatorname{Tr}\Bigl[\bigl( {N}_{2}^{[2]}\bigr)^2\Bigr] - \operatorname{Tr}\Bigl[\bigl(
{N}_{2}^{[2]}\bigr)^3\Bigr]\right)\label{eq:phi2}.
\end{align}
The trace of a matrix $X^{[2]}$ is denoted as $\operatorname{Tr}[X^{[2]}]$. The $\phi_2$ term was introduced by Tarazona \cite{Tarazona2000FMT} within a
dimensional crossover analysis.\cite{FMT2} In the edFMT \mbox{$c_0 = 3/(16\pi)$} was chosen. In this way the exact third virial coefficient of hard spheres is
included. For general convex rotating (\emph{i.e.} non-parallel) particles, the edFMT is exact up to the second virial coefficient for the isotropic fluid.

\subsection{edFMT of parallel hard spherocubes}

We now apply the edFMT to a monocomponent system of parallel hard spherocubes. We chose the model of spherocubes as the curvatures of all 
sections of its surface are
\textit{constant}, where we divided the surface of the particle into its spherical, cylindrical and flat components, see Fig.~\ref{fig:shape}.
It is thus sensible to split the weight functions into a sum of contributions, each of which is related to one of these sections of the surface
or, for $n_3$, to the corresponding volume.
Since the convolution is a bilinear operation, the weighted densities also decompose into a sum of terms related to the different components.
Due to this decomposition we are able to determine the weighted densities for each component separately and use a coordinate system
appropriate for its geometry.

\subsubsection{Homogeneous fluid}

As the most simple case, we first study the monocomponent homogeneous fluid. It is characterized by a constant density profile $\rho(\mathbf{r}) \equiv N/V$. Consequently,
the weighted densities $\{n_\alpha\}$ are independent of the position vector and read $n_\alpha = \rho m_\alpha$, with $m_\alpha = \int \!\!\mathrm{d}{\mathbf{r}}
w_{\alpha}$. The integrated scalar weight functions $m_\alpha$ for $\alpha=3,2,1,0$ represent the volume, surface, mean half width~\cite{Mulder2005spherozonotopes} and Euler characteristic of a spherocube.
Furthermore, we notice that the packing fraction $\eta$ is equal to $n_3$. Thus, we can express the weighted densities in dependence of the packing fraction as
\begin{align*}
	n_\alpha = \eta \frac{m_\alpha}{m_3}.
\end{align*}
In the case of the scalar weighted densities we obtain
\begin{align*}
 m_{0} &= 1 \\
 m_{1} &= \tfrac{1}{4}\left(3l - d\right)\\
 m_{2} &=  \pi d^2 + 3\pi d \sigma + 6\sigma^2\\
 m_{3} &=  v_\text{rc}.
\end{align*}
The $\mathbf{n}_2$ vector--type weighted density vanishes for the homogeneous fluid, which is a consequence of the Gauss' divergence theorem and holds for arbitrarily
shaped particles.\cite{Hansen-Goos2010edFMTlong} As a result, the vector term in the free energy density~(\ref{eq:free_energy_density}) vanishes. Additionally,
the tensor weighted density $\smash{{N}_{1}^{[2]}}$ is zero due to the cubic symmetry of the particle and the traceless nature of $\smash{N_1^{[2]}}$. 
As a result, truncation at $J=2$ order would leave us only with the scalar terms, which does not result in the correct second virial coefficient.
To improve the theory, we need to extend \eqref{eq:Mayer} at least to the first 
tensor term that does not vanish in the homogeneous fluid. In the case of spherocubes, the first nonzero term contains tensors of fourth order, i.e. $J=4$.
The corresponding generalized product reads
\begin{align*}
{N}_{1}^{[4]} \star {N}_{2}^{[4]} = \frac{\eta^2}{m_3^2} \left( \frac{3}{8}\sigma^3 + \frac{9}{128}\pi d \sigma^2 \right).
\end{align*}
In this way we can determine the free parameter $\zeta_4$ by comparison with the exact second virial coefficient $B_2$. The virial expansion up to the third
virial coefficient $B_3$ reads
\begin{align}\label{eq:virial}
	\frac{P}{\rho k_\mathrm{B} T} = 1 + \frac{B_2}{v_{\text{rc}}} \eta + \frac{B_3}{v_{\text{rc}}^2} \eta^2 + \mathcal{O}(\eta^3),
\end{align}
with $B_2 = 4 v_{\text{rc}}$ for spherocubes as can be shown by elementary geometrical considerations. On the other hand, the compressibility factor is given by the derivation of the free energy density with respect to $n_3$:
\begin{align}\label{eq:PhiVirial}
	\frac{P}{\rho k_\mathrm{B} T} &= \frac{\eta}{N}\frac{\partial \mathcal{F}}{\partial \eta} = \frac{1 + a(\zeta_4)\eta + b(\zeta_4,c_0)\eta^2}{(1-\eta)^3},
\end{align}
where 
\begin{align*}
a(\zeta_4) &= \frac{m_1 m_2 - \zeta_4 {M}_{1}^{[4]} \star {M}_{2}^{[4]}}{m_3} - 2,\\
b(\zeta_4,c_0) &= c_0\frac{4m_2^3 }{9 m_3^2} - a(\zeta_4)-1. 
\end{align*}
The Taylor expansion of Eq. \eqref{eq:PhiVirial} around $\eta = 0$ is
\begin{align}\label{eq:virfit}
\frac{P}{\rho k_\mathrm{B} T}	
= 1 + (a \! + \! 3)\eta + (3a \! + \! b \! + \! 6)\eta^2 + \mathcal{O}(\eta^3),
\end{align}
where we dropped the $\zeta_4$ and $c_0$ dependences for brevity.
By comparing the terms of order $\eta$ in~\eqref{eq:PhiVirial} and~\eqref{eq:virial} with each other, 
we obtain $a(\zeta_4) = 1$. Solving this equation for $\zeta_4$ yields
\begin{align}
\label{zeta}
\zeta_4 &= \frac{m_1 m_2 - 3 m_3}{{M}_{1}^{[4]} \star {M}_{2}^{[4]}}.
\end{align}
The free parameter $\zeta_4$ only depends on the shape parameter $s$.

In the limit of hard parallel cubes ($s=0$), we do not recover the equation of state found by Cuesta and Mart{\'i}nez--Rat{\'o}n,\cite{martinez-raton1999}
because for non--spherical shapes the third virial coefficient is not exact. We can resolve this issue by adjusting the constant $c_0$, which was introduced in \eqref{eq:phi2}.
Equating the third virial coefficient in \eqref{eq:virial} with the one in \eqref{eq:virfit} and inserting $a(\zeta_4)=1$,
yields $b(\zeta_4,c_0) = B_3v_{\text{rc}}^{-2} - 9$, which is equivalent to
\begin{align*}
c_0 = \frac{9 m_3^2}{4 m_2^3}\Bigl(\frac{B_3}{v_{\text{rc}}^2} - 7\Bigr).
\end{align*}
In contrast to $B_2/v_\text{rc}=4$, the dimensionless third virial coefficient $B_3/v_\text{rc}^2$ depends on the shape parameter and has to be determined for
every $s$. 
We have numerically calculated the third virial coefficient using Monte--Carlo integration with an approximate error of $\approx
10^{-4}$. The result $B_3(s)$ smoothly interpolates between the analytically known $B_3$ for cubes \mbox{$B_{3}(0) = 9v_{\text{rc}}^2$}~\cite{PercusYevick1958}
and spheres \mbox{$B_{3}(1) = 10v_{\text{rc}}^2$}.\cite{Zwanzig_cube_gas}
Using the respective third virial coefficients of these limiting cases, we recover the FMT equations of state of spheres and parallel hard cubes.

\subsubsection{Simple cubic crystal}

We parametrize the density profile of the crystal by a standard Gaussian form given by
\begin{align*}
 \rho(\alpha,\nu_\text{vac},\mathbf{r}) = (1-\nu_\text{vac}) \left( \frac{\alpha}{\pi} \right)^{\frac{3}{2}} \sum_{\mathbf{R}} \exp(-\alpha(\mathbf{r} - \mathbf{R})^2),
\end{align*}
where $\{\mathbf{R}\}$ are the lattice vectors of the prescribed crystal structure. We regard \mbox{$\nu_\text{vac} \in [0, 1]$} as the vacancy concentration and \mbox{$\alpha \in [0,\infty)$} as the Gaussian parameter that characterizes the profile. For a simple cubic crystal structure the parametrization factorizes and takes a simpler form with lattice constant $a_0 = ((1-\nu_\text{vac}) m_3/\eta)^{\frac{1}{3}}$, provided that $(1-\nu_\text{vac}) m_3 \geq \eta l^3$. With this parametrization we can determine the weighted densities according to \eqref{eq:ndef}.
As for the homogeneous fluid, we truncate the tensor expansion at $J=4$.
For inhomogeneous density distributions, the generalized products of the second and third order tensors in general do not vanish. Thus, we need to determine
the free parameters $\zeta_2$ and $\zeta_3$. In the special case $s=0$, the infinite tensor expansion in Eq.~(\ref{eq:phi1}) with $J=\infty$ and
$\zeta_j=(-1)^j$ can be evaluated analytically for the simple cubic crystal, 
which gives the same result as the truncated $\phi_1$ in Eq.~\ref{eq:phi1} with $(\zeta_2,\zeta_3,\zeta_4) = (1,-3,4)$ for $J=4$. In this work, we require that the truncated free
energy functional is identical to this analytical free energy functional for the simple cubic crystal in the limit $s \rightarrow 0$. Accordingly, we set
$(\zeta_2,\zeta_3,\zeta_4) = (1,-3,\zeta_4)$ for spherocubes with finite $s$, where $\zeta_4$ is given by the value for the homogeneous fluid \eqref{zeta}.
Numerical minimization of the free energy functional with respect to $\alpha$ and $\nu_\text{vac}$ yields a continuous freezing transition  
for $s\leq 0.65$.

\begin{figure*}[t!]
\includegraphics[width=0.85\textwidth]{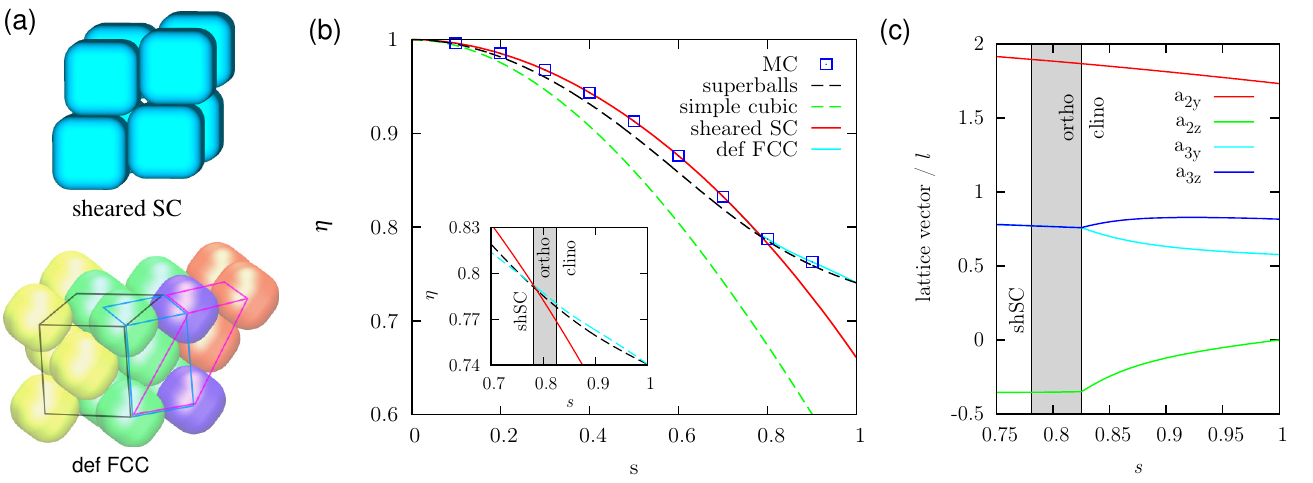}

\caption{
(a) The crystals that were found using the unit cell simulations: a "sheared" version of a simple cubic crystal (shSC) and 
a deformed face-centered cubic crystal (def FCC). For the deformed FCC crystal, the deformed cubic unit cell of FCC is indicated in black (yellow and green particles), the 
body centered orthorhombic (ortho) unit cell in blue (green and blue particles) and the  base centered
monoclinic (clino) unit cell in magenta (red, blue and some of the green particles). (b) The packing fractions of the various crystals as a function of the aspect ratio $s$ at close packing.
For comparison, the packing fraction at close packing for superballs which are mapped onto spherocube with aspect ratio $s$ (see Appendix~\ref{sec:super_comp}) are
included, as well as the packing fraction of the simple cubic phase. The inset shows an enlargement of the region where the def FCC phase has the highest
packing. (c) The lattice vectors of the ortho and clino variants of def FCC. In (b) and the inset of (c) the region where the various phases have the 
highest packing are indicated by the labels ``shSC'', ``ortho'' and ``clino'' and the gray area.
\label{fig:packing}}
\end{figure*}

\section{Results} \label{sec:results}

\subsection{Crystals and regular close packing}\label{sec:packing}

We found two different types of crystals in the unit cell simulations described in Sec.~\ref{subsec:uc_sims}. The crystal found at low aspect ratios
(for near cubes) resembles a sheared version of the simple cubic phase, see~Fig.~\ref{fig:packing}(a). The (primitive) lattice vectors at close packing are given by
\begin{equation}
\mathbf{a}_1 = \begin{pmatrix} l \\ \Delta a \\ \Delta a \end{pmatrix},\qquad
\mathbf{a}_2 = \begin{pmatrix} \Delta a \\ l \\ \Delta a \end{pmatrix},\text{ and }
\mathbf{a}_3 = \begin{pmatrix} \Delta a  \\ \Delta a \\ l \end{pmatrix},
\end{equation}
where $\Delta a$ is given by $\Delta a\equiv d$\hspace{0.3pt}\raisebox{1.0pt}[0pt][0pt]{$\big($}\hspace{0.3pt}$1-1/\sqrt{2}$\hspace{1.5pt}\raisebox{1.0pt}[0pt][0pt]{$\big)$}.
The Bravais lattice of the sheared simple cubic phase (shSC) is the rhombohedral lattice.
At close packing, the packing fraction of the shSC crystal is 
\begin{equation}
\eta_\text{shSC}=v_\text{rc}
/ \left\{ l^3 - 3 \Delta a^2 l + 2 \Delta a^3 \right\}.
\end{equation}

The shSC crystal has the highest packing for $s<0.781133$; for higher aspect ratios,
a phase similar to face centered cubic was encountered, which we called deformed FCC (def FCC), also depicted in Fig.~\ref{fig:packing}(a). This phase
actually consists of two phases between which a continuous transition is observed.
Decreasing $s$ from 1 (that is starting with spheres), FCC is deformed such that only a base centered monoclinic (BCM) unit cell can be recognized
(a BCM unit cell can also found in the cubic unit cell of FCC). As the aspect ratio is decreased beyond 0.825079, a body-centered orthorhombic unit cell
is found to have the highest packing fraction. Both crystals can be described using the BCM unit cell (which is the most general):
\begin{equation}
\mathbf{a}_1 = \begin{pmatrix} l \\ 0 \\ 0 \end{pmatrix},\qquad
\mathbf{a}_2 = \begin{pmatrix} 0  \\ a_{2y} \\ a_{2z} \end{pmatrix},\text{ and }
\mathbf{a}_3 = \begin{pmatrix} 0  \\ a_{3y} \\ a_{3z} \end{pmatrix},
\end{equation}
where $a_{i\nu}$ for $i=2,3$ and $\nu=y,z$ are to be determined and the `base' of the unit cell,
on which both particles in the unit cell lie, is spanned by $\mathbf{a}_1$ and $\mathbf{a}_2$.
The components $a_{i\nu}$ in general have to be calculated numerically. The ones at close packing are plotted in Fig.~\ref{fig:packing}(c).

The packing fraction at close packing for these crystals are shown in Fig.~\ref{fig:packing}(b). For comparison, the close-packed simple cubic
crystal is also shown. Clearly, the simple cubic phase has a lower maximal packing fraction than the sheared simple cubic phase for $s>0$
and, naively, one would think that the sheared simple cubic crystal is more stable for all densities. However, we will show below that the fluid first transforms into
a simple cubic crystal phase as the pressure is increased for a large range of $s$ values. This shows once again that packing arguments
should not be used to infer the stable crystal at finite pressures.
We also included the maximal packing fraction for superballs in Fig.~\ref{fig:packing}(b) as a function of
the $s$-value of the spherocube that has a minimal Hausdorff distance to the superball, see Appendix~\ref{sec:super_comp}. Cube-like superballs show two distinct crystal phases at close
packing,\cite{Jiao_SB_pack}
which are quite similar to our shSC and def FCC phases. Clearly, superballs have a lower packing fraction at close packing
for all values of $s$, which is especially marked at low $s$, where the flat faces of the spherocubes allow a very efficient packing into the
shSC phase.

\subsection{Comparison between FMT and simulations}

\begin{figure}[t!]
\includegraphics[width=0.4\textwidth]{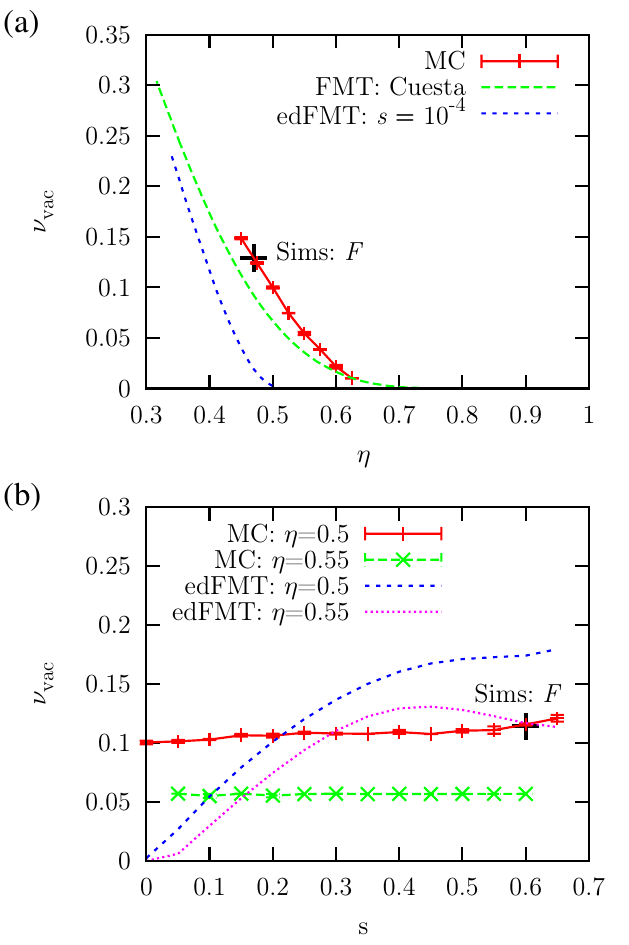}
\caption{
(a) The vacancy concentration $\nu_\text{vac}$ as a function of packing fraction $\eta$ as obtained from variable box edge length $NVT$ simulations for
 parallel hard cubes, that is with $s=0$, 
  from Cuesta \emph{et al}'s FMT~\cite{Cuesta_paraPRL1} for parallel hard cubes and the edFMT of this work for rounded cubes with
$s=10^{-4}$. (b) The same as (a), but now $s$ varies and $\eta$ is fixed to $0.5$ or $0.55$. 
The black points is determined by minimizing the free energy with respect to the vacancy concentration for $s=0$ and $\eta=0.47$ in (a) and for $s=0.6$ and
$\eta=0.5$ in (b), see Fig.~\ref{fig:f_of_vac}.
\label{fig:f_vac}} 
\end{figure}

In this section we compare the data from the Monte Carlo and event-driven MD simulations to FMT results for the simple cubic crystal.
In Figs.~\ref{fig:f_vac}(a) and (b), 
we show the vacancy concentration as measured in variable box length $NVT$ Monte Carlo simulations
and compare with the results from FMT. Also shown are two black points, which are determined from free energy calculations.
Reassuringly, these points correspond well to the other simulation results.
 Fig.~\ref{fig:f_vac}(a) shows the dependence on the packing fraction for fixed aspect ratio $s=0$.
The trend of the vacancy concentration from edFMT corresponds to that of the MC simulations, as does the original FMT for hard parallel
cubes by Cuesta~\emph{et al}\cite{Cuesta_paraPRL1}. However, both theories underestimate the vacancy concentration considerably. Possibly, fluctuations, which are 
absent in the theory, stabilize crystals with higher vacancy concentrations. The dependence of the vacancy concentration on the aspect ratio
$s$ is shown in Fig.~\ref{fig:f_vac}(b). The functional by Cuesta \emph{et al}~\cite{Cuesta_paraPRL1} cannot be applied for rounded hard cubes with $s\neq 0$. While the simulation data shows only a very weak dependence on $s$, the theoretical result 
increases dramatically with increasing $s$, such it actually \emph{over}estimates the data at $s\gtrsim 0.1$. Nevertheless, both theory and
simulations show that the high vacancy concentration is not an artifact of the sharp edges of the parallel hard cubes, but remains
also when the edges are rounded. 

\begin{figure}
\includegraphics[width=0.4\textwidth]{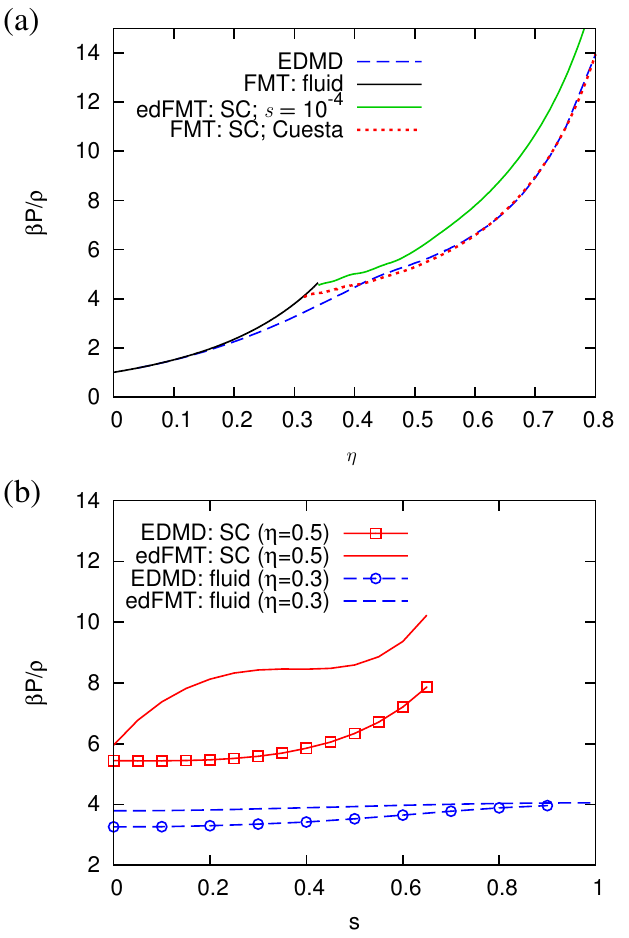}
\caption{
(a) The compressibility factor $P/\rho k_B T$ as a function of packing fraction $\eta$ as obtained from event-driven MD simulations for
 parallel hard cubes ($s=0$), from Cuesta \emph{et al}'s FMT~\cite{Cuesta_paraPRL1} for parallel hard cubes and the edFMT of this work for rounded cubes with
$s=10^{-4}$. The FMT and edFMT results for the fluid are exactly equal. (b) The same as (a), but now $s$ varies and $\eta$ is fixed to $0.3$, where the fluid is found, or $0.5$, where the simple cubic
crystal is stable.
\label{fig:eos}} 
\end{figure}

In Fig.~\ref{fig:eos}, the equation of state ($P/\rho k_B T$) in the homogeneous fluid and in the simple cubic crystal with the vacancy concentration of Fig.~\ref{fig:f_vac} is shown. 
Again, simulation results are compared to FMT results obtained using the functional from this work and from Ref.~\onlinecite{Cuesta_paraPRL1}.
The agreement with the simulation data is reasonable for $s\simeq 0$, while the previous functional~\cite{Cuesta_paraPRL1} for parallel hard cubes
describes the simulation data the best.
As $s$ increases, the agreement between theory and simulations deteriorates somewhat, and for large values of $s$ improves again, such that the
difference in the compressibility factor resulting from the theory and the simulation is of the order of 1 for $s\gtrsim 0.1$.

\begin{figure}[t!]
\includegraphics[width=0.4\textwidth]{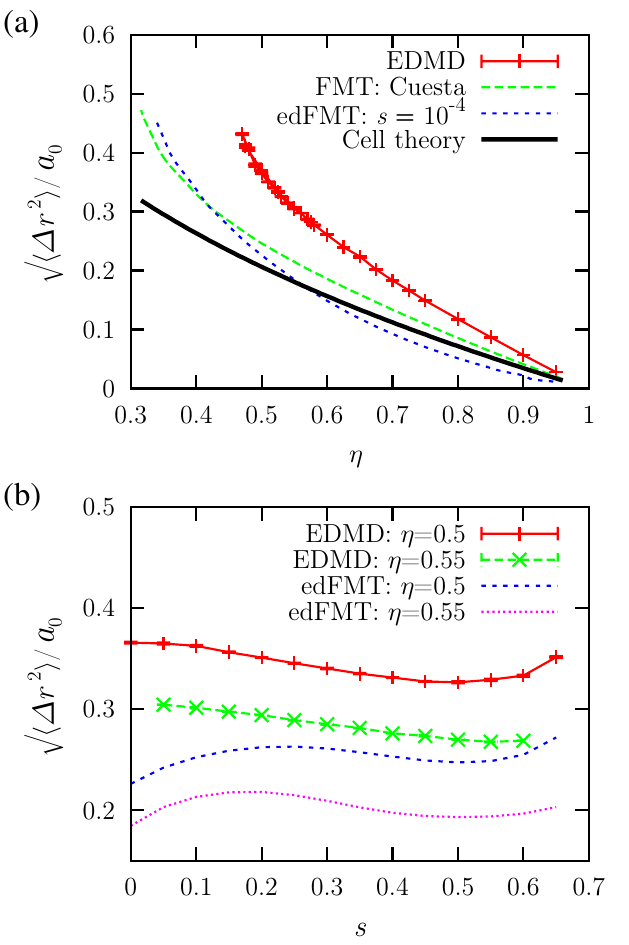}
\caption{
The dimensionless root mean squared deviation (RMSD)  from the nearest lattice site divided by the lattice constant, $\Delta r/a_0$, as obtained from variable
box edge length $NVT$
simulations. Results from our edFMT and the older FMT~\cite{Cuesta_paraPRL2} are also shown.  (a) The RMSD as a function of packing fraction $\eta$ for 
 parallel hard cubes, that is, for $s=0$. The thick black line denotes the cell theory result without
vacancies. (b) The RMSD for varying $s$ at $\eta=0.5$ and $\eta=0.55$.
\label{fig:rmsd}} 
\vspace*{-1em}
\end{figure}
The root mean squared deviation from the nearest lattice site (RMSD) as measured in event-driven molecular dynamics simulations (EDMD) is compared to FMT results 
in Figs.~\ref{fig:rmsd}(a) and (b).
Again, the first of these figures shows the $\eta$ dependence at $s=0$,
while the second shows the $s$ dependence at fixed $\eta$. The RMSD is often compared with the Lindemann criterion~\cite{Lindemann_crit} for first order phase transitions, which says
that the crystal starts to melts when the RMSD is around 10\%-20\% of the lattice constant.
As expected for second order phase transitions, the RMSD of the simple cubic crystal of (rounded) cubes
is higher than the Lindemann parameter at the transition; it is in fact 
two to four times as high.
The theoretical RMSD results at $s=0$ [Fig.~\ref{fig:rmsd}(a)] again show the correct trend, but both  
theories under-estimate the simulation results, as was the case with the vacancy concentration. The result from cell theory (for zero vacancies) is also
indicated by the think line. Interestingly, the simulation results have a substantially different slope than cell theory even when approaching close packing,
the MSD from simulations is approximately 1.6 times the cell theory result.
For comparison, the mean squared displacement measured in simulation of hard spheres is approximately 1.098(4)~\cite{Young_Alder_MSD} times the cell theory result. 
The dependence of the RMSD on the $s$ of FMT in Fig.~\ref{fig:rmsd}(b) is qualitatively very similar to the simulation results: At $\eta=0.55$ the RMSD
decreases monotonically,
while the RMSD for $\eta=0.55$ shows a strong decrease with increasing $s$ for small $s$ followed by a small increase when $s$ is increased beyond a certain
value $s\simeq 0.55$.

\subsection{Phase diagram}

\begin{figure*}[t!]
\centering\includegraphics[width=0.7\textwidth]{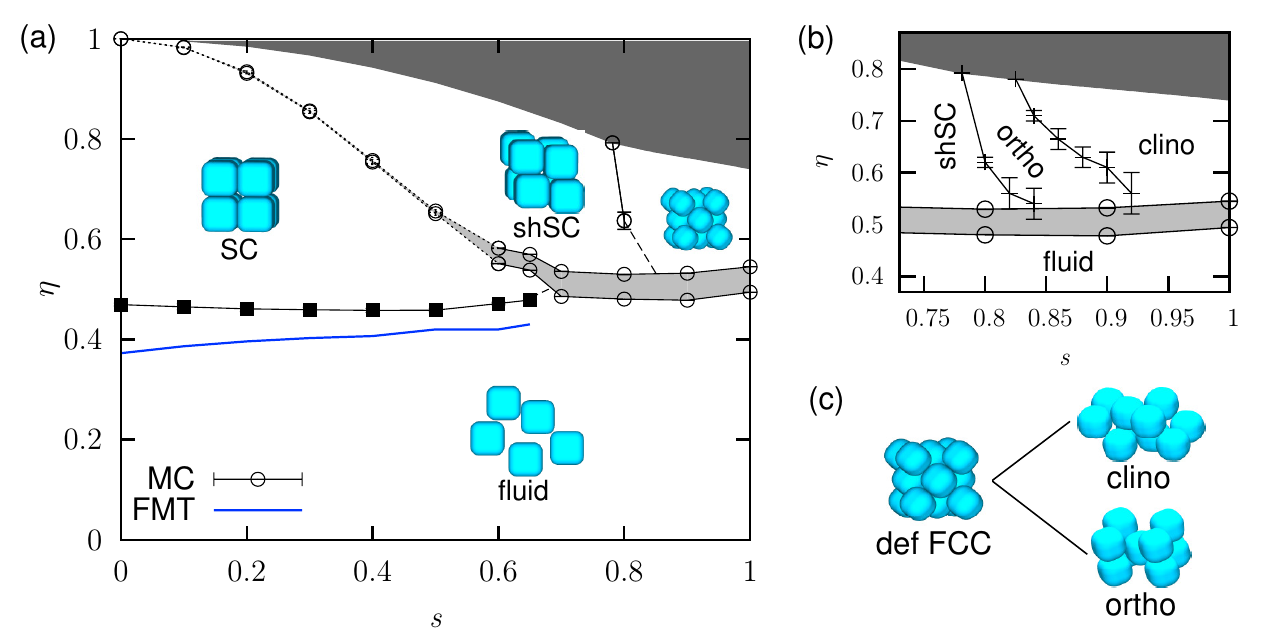}

\caption{
(a) The phase diagram of parallel rounded cubes in the $\eta$--$s$ representation, where  $\eta=v_\text{rc} N/V$ is the packing fraction with $v_\text{rc}$
the volume of a rounded cube and  $s=d/l$ is the rounding parameter (see Fig.~\ref{fig:shape}).
A cube has $s=0$ and a sphere $s=1$. Shown are the areas of stability of the deformed $fcc$ phase
of near spheres (def FCC), the sheared cubic crystal (shSC), the simple cubic crystal (SC) and the fluid phase in white. The forbidden region above the 
close packing density is shown in dark gray and coexistence areas in lighter gray (coexistence lines are vertical). 
The filled symbols (MC simulations) and the thick line (FMT) denote second order phase transitions, while the empty symbols denote 
first order phase transitions from simulations.
(b) An enlargement of the large $s$ region of the phase diagram: the def FCC phase is actually seen to have a body-centered orthorhombic variant (ortho)
and a base-centered monoclinic variant (clino), as depicted in (c).
\label{fig:ph-dia}}
\end{figure*}

The phase diagram of parallel hard rounded cubes is shown in Fig.~\ref{fig:ph-dia}. 
The fluid to simple cubic (SC) crystal second order transitions for hard parallel cubes $s=0$ from this work 
have a critical packing fraction, $\eta_c=0.469(3)$ from simulations and $\eta_c=0.3325$ from FMT, which should be compared
to the earlier simulation~\cite{groh2001cubes} critical density, $\eta_c=0.53(1)$, and the result from the original FMT,\cite{martinez-raton1999} $\eta=0.3143$.
Our simulation results differ from the previous work because it was assumed that the SC crystal had zero vacancies in Ref.~\onlinecite{groh2001cubes}.
In contrast, we find an extremely high vacancy concentration of $13\%$ at coexistence. The FMT vacancy concentration of the earlier
FMT~\cite{martinez-raton1999} was 30\%, while our edFMT gives 23\%. Note, that the critical densities differ for the two theories and the simulations,
which explains the reversal of the trends compared to the results at fixed packing fraction. Our FMT describes 
our simulation results slightly better than the earlier FMT,\cite{martinez-raton1999} as far as the critical density and the vacancy concentration 
at $\eta_c$ are concerned.
Conversely, the inclusion of vacancies has brought the critical density from simulations closer to the FMT results.

For hard parallel cubes with a finite rounding (\emph{i.e.} $s\neq 0$), the transition from the fluid to the SC crystal phase is also
second order both for the simulations and for FMT. For the simulations, this is indicated by the critical scaling whose exponents belong to the Heisenberg
universality class.\cite{groh2001cubes} The reasonable agreement between the simulation results and the FMT at $s=0$ is even slightly improved for $s\neq 0$, see
Fig.~\ref{fig:ph-dia}. The phases that were found using the simulations of single unit cells all have their separate area of stability in the phase 
diagram. At low $s$, the SC is stable at low densities, while the sheared variant (shSC) is stable at high densities. 
The transition from SC to shSC seems to become more weakly first order as $s$ approaches zero, and simultaneously the vacancy concentration
decreases. For $s\leq 0.5$, we did not use free energy calculations, because the free energies of SC and shSC were very close making it hard
to find the transition. Instead, we used direct simulations, which always lead to a pressure at which the difference in chemical potential
was very small.
As $s$ is increased, the SC--shSC
transition goes down in density and at some point the shSC coexist directly with the fluid. Finally, the deformed FCC phase is stable for sphere-like 
particles ($s\geq 0.8$). As mentioned in Sec.~\ref{sec:packing}, the def FCC phase actually consists of two crystals, one with a base-centered monoclinic
(clino) unit
cell and an other with a body-centered orthorhombic (ortho) unit cell. 
Our motivation for investigating the clino to ortho transition in this system, is that
pyroxene, 
the second most abundant mineral in the earth's mantle, also has clino and ortho forms.\cite{Ulmer_clino_ortho}
As a result, 
the clino--ortho transition of pyroxene is a topic of great interest in geology.
In our case, the ortho-unit cell needs only be slightly deformed to form the clino unit cell: the angle between two of the orthorhombic lattice vectors 
is changed to slightly to a little more or less than 90 degrees; the difference is at maximum 2.22 degrees for $s=0.884$ at close packing.
The packing fractions at the transition from clino to ortho and the transition from shSC to ortho were obtained by direct simulations, as shown in Fig.~\ref{fig:ph-dia}.
The shSC--ortho phase transition is more strongly first order than the ortho--clino transition, which enabled us to calculate
the free energy of the shSC and ortho phases separately and, reassuringly, the free energy difference between the two phases at the shSC--ortho transition, which was obtained in direct
simulations, is smaller than the statistical error.

\section{Conclusions}  \label{sec:conclusions}

We studied a system of parallel rounded cubes (spherocubes) using fundamental measure theory and simulation with
a special emphasis to the second order freezing into the simple cubic phase. 
We developed the fundamental measure theory starting from edFMT~\cite{Hansen-Goos2009edFMTPRL,Hansen-Goos2010edFMTlong} expanding
up to fourth order tensor terms and renormalizing the third and fourth order terms.
 When we apply the theory to
the simple cubic phase, we find that the freezing is second order, not just for perfect cubes
with shape parameter $s=0$, but also when we introduce a degree of rounding $s$ up to $s=0.65$. 
This finding is confirmed by finite size scaling techniques using Monte Carlo (MC) simulations.
Furthermore, we find both in theory and simulations an unusually high vacancy concentration, namely 13\%, that is, about twice as high 
as for rotating perfect cubes~\cite{Smallenburg_MM_cubes} and four orders of magnitude higher than that of hard spheres.\cite{Bennett_vacancies,Oettel_Schilling_MC_FMT_HS} 
The very high vacancy concentration and the simple overlap criterion make this system an ideal system to study vacancies
(the number of vacancies can always be decreased by increasing the density if so required). 

When comparing the theory to the simulations, we find good qualitative agreement; exceptions are
the dependence of the vacancy concentration and the pressure on the shape parameter $s$ which show the incorrect trend for FMT.
Quantitative differences between the FMT and simulation results are found for most quantities. This is most likely
caused by anomalous higher virial coefficients for this system (parallel cubes for instance have negative sixth and seventh virial
coefficients~\cite{Hoover_vir_cube}) which are not reproduced by the theory. Nevertheless,
the most important property, the packing fraction at freezing is predicted quite well by FMT over the whole range of $s$
where the simple cubic crystal is stable. 

Finally, we completed the phase diagram of rounded cubes by investigating the possibility of other crystal phases
in direct simulations and by performing free energy calculations 
using the results obtained in these simulations.
The phase diagram of parallel rounded cubes is surprisingly rich considering the simplicity of the model:
it contains, apart from the simple cubic crystal phase, three more crystal phases. For low values of $s$ and
high densities a sheared variant of the simple cubic (shSC) is found, while at low densities the simple cubic crystal is stable. For higher values of $s$,
first the simple cubic crystal and later also the shSC phase disappears to be replaced by a body-centered orthorhombic crystal which is essentially a slightly deformed face-centered
cubic crystal. Finally, a base-centered monoclinic crystal is found for values of $s$ near one, that is, for near spheres.
The resulting symmetry change is interesting because a similar transition is found for an abundant mineral in the earth's mantle.\cite{Ulmer_clino_ortho}
We expect that most, if not all, of the crystal phases we observed can also be found for rotating rounded cubes, which could be verified in experiments on colloidal rounded cubes.
\acknowledgments

We thank Klaus  Mecke, Marjolein Dijkstra, Laura Filion and Frank Smallenburg for useful discussions.
This work was financially supported by the DFG within SFB TR6 (project D3).

\appendix 

\section{Overlap algorithm and collision prediction} \label{sec:overlap_and_collision}

The criterion for overlap between two (co-aligned) spherocubes is surprisingly simple:
Two spherocubes 
overlap when shortest distance, $\Delta r_{ij}$, between any point on the surface of particle $i$ and any point on $j$ is smaller than $d$.
The shortest distance can be calculated in the following two steps:
\begin{align}
b_{ij,\nu}&\equiv|r_{j,\nu}-r_{i,\nu}|-\sigma \label{eqn:def_bijalpha}\\[0.8em]
\Delta r_{ij,\nu} & = \left\{\begin{array}{cc} \mathrm{sign}(r_{j,\nu}-r_{i,\nu})b_{ij,\nu}\, & b_{ij,\nu}\geq 0 \\[0.5em] 0
&\text{otherwise}\end{array}\right. \label{eqn:Deltarij}
\end{align}
the norm of the vector with components $\Delta r_{ij,\nu}$ for $\nu=x,y,z$ is the shortest distance $\Delta r_{ij}$.
The simplicity of the overlap criterion for spherocubes is a large advantage compared to 
superballs for which it can not be ascertained using analytical means whether two particles overlap or not.\cite{Batten}
Note, that the overlap criterion for spherocubes becomes more complicated when the
particles are not
aligned with the Cartesian axes.

Collisions can be analytically predicted for parallel hard rounded cubes as follows:
The surface of a rounded cube consists of sections of axis-aligned cylinders, planes and spheres, for which collisions
can be easily calculated by the collision detection algorithms for the corresponding one, two and three dimensional hyper-spheres.\cite{Alder_EDMD} 
Specifically, a collision test for two $n$-dimensional hyperspheres is required when exactly $n$ components of $\mathbf{b}_{ij}$ [the vector with components 
$b_{ij,\nu}$, see Eq.(~\ref{eqn:def_bijalpha}) ]
are nonzero (at least one of components of $\mathbf{b}_{ij}$ is non-zero initially in an overlap-free configuration).
The time at which a certain component of $\mathbf{b}_{ij}$ becomes nonzero can easily calculated from Eq.~(\ref{eqn:def_bijalpha}).
All such times are determined, sorted and inserted in a list, whose subsequent elements define the time intervals at 
which certain parts of the surface of one of the particles might collide with a part of
the other particle's surface. For each such a time interval, the corresponding hypersphere collision check is performed 
and the shortest of the resulting times is the time of collision of the two spherocubes.
The rest of the algorithm is the same as the optimized algorithm for hard spheres,\cite{Rapaport_bin_tree} in which the collisions are stored, together with others event (such
as measurements) in a binary tree leading
to a theoretical $N \log(N)$ scaling of the computational effort for a fixed run time.\cite{Rapaport_bin_tree} 

\section{The Frenkel-Ladd method for crystals with vacancies} \label{sec:FrLadd_vac}

In the original Frenkel-Ladd~\cite{frenkel1984einstein} approach, each particle is coupled to its ideal lattice position with a harmonic spring, such that
the external coupling potential reads: 
\begin{equation} \beta U_\text{har}({\bf r}^N;\lambda) = \lambda
\sum_{i=1}^{N} ({\bf r}_i-{\bf r}_{0,i})^2/l^2,
\end{equation} where  ${\bf r}_i$ denotes the position
of particle $i$ and ${\bf r}_{0,i}$ the
lattice site of particle $i$ and $\beta=1/k_BT$.
If the value of $\lambda$ is high enough or if the lattice positions are far enough apart, the particles do not interact and 
the free energy of the system is given by the known analytical free energy of the non-interacting Einstein crystal.\cite{frenkel1984einstein}
Therefore, the coupling constant $\lambda$ can be used to switch between an ideal Einstein crystal for high $\lambda$ and 
the unperturbed crystal for $\lambda=0$. The free energy of the crystal for $\lambda=0$ can then be found by integrating over $\lambda$:
\begin{equation}
 f^*(N,V,T)=f^*_\mathrm{Einst}(N,V,T)
 -\int^{\lambda_\mathrm{max}}_0 \mathrm{d}\lambda  \left\langle
\frac{\partial f^*}{\partial \lambda}
\right\rangle 
\end{equation}
where $\langle \partial f^*/\partial \lambda 
\rangle\!=\!\big\langle U_\text{har}({\bf r}^N;\lambda)\rangle/(\lambda N)$.
For $\lambda=0$, $\langle U\rangle$ diverges as the center of mass of the system diffuses as a whole, taking the particles ever further away
from their lattice position. To overcome this problem, the center of mass is fixed which results in additional (small) terms in the free energy of
the non-interacting system, which can be found in Refs.~\cite{frenkel1984einstein,FrenkelSmit}.
The value of $\lambda_\text{max}$ required to obtain a non-interacting Einstein crystal depends on the lattice spacing, such that 
free energy calculations over a wide range of densities require constant tuning of $\lambda_\text{max}$.
However, the same value for $\lambda_\text{max}$ can be used for every density
if the inter-particle potential is replaced by a purely repulsive finite potential whose interaction strength
is slowly decreased from essentially infinite to zero (where an essentially infinite interaction strength implies that no overlap is
found during the simulation).
The free energy difference between the interacting crystal  
for $\lambda=\lambda_\text{max}$ and the non-interacting Einstein crystal is then obtained by integrating over the strength of interaction $\gamma$ of
the inter-particle potential, see Refs.~\cite{fortini_cryst_conf,MM_Dijkstra2008db} for details. The soft interaction between two particles, in this case, reads $\gamma(1-0.9 \Delta
r_{ij}/l)$, see Eq.~(\ref{eqn:Deltarij}). We have used this method for all crystal phase with exception of the simple cubic crystal phase which
had a large concentration of vacancies.

When the crystal has a nonzero vacancy concentration, the Frenkel-Ladd method~\cite{frenkel1984einstein} requires some modifications. First of all, the particles are no longer
associated with a single lattice site as they can hop to a different site when it is empty. Therefore, the harmonic potential can no longer be used.
Instead, we use the periodic external potential proposed by Groh and Mulder,\cite{groh2001cubes} which reads
\begin{equation}
\beta U_\text{per}(\mathbf{r}^N)=\lambda \sum_{i=1}^N \sum_{\nu=x,y,z} 1-\cos(k_\nu r_{i,\nu}), \label{eqn:Uper}
\end{equation}
where $k_\nu=2\pi N_\nu/L_\nu$. The number of unit cells $N_\nu$ in direction $\nu$ and the length $L_\nu$ of the edge of the box are adjusted 
to tune the density and vacancy concentration. Furthermore, we promote hopping of a particle from a filled lattice site to an empty one also at large
$\lambda$ by performing moves of exactly one (cubic) lattice vector, which allows the distribution of vacancies over the lattice to equilibrate.
Note, that the center of mass of the system is already fixed by the external potential, so no additional terms due to
the fixing of the center of mass arise in the free energy.
However, additional moves which translate the whole system homogeneously are required 
to efficiently equilibrate the center of mass.
Finally, the ideal Einstein crystal free energy itself is modified because of the modified inter-particle potential.
Additionally, the combinatorial free energy of choosing $N$ filled lattice sites
out of a total of $M$ lattice sites needs to be included in the free energy. The total free energy of the non-interacting Einstein crystal reads
\begin{equation}
f^*_\text{Einst}=-\frac{1}{N}\ln\left[\frac{M!}{(M-N)! N!}\right]-3 \ln z_1(\lambda_\text{max}) 
\end{equation}
where  $z_1(\lambda)=I_0(\lambda) \exp(-\lambda)\, a_0/\Lambda$ is the factor in the partition
sum which results from the integration of the degrees of freedom of a single particle in one direction, 
and $I_0$ is the zeroth modified Bessel function of the first kind. 
If $\lambda$ goes to infinity, the external potential (\ref{eqn:Uper}) can approximated by a harmonic potential,
and simultaneously $z_1(\lambda)$ approaches $\sqrt{\lambda/2\pi}\, a_0/\Lambda$
as $\lambda_\text{max}\to\infty$. 
It is this expression for $z_1(\lambda_\text{max})$, which we use in practice, because the approximation has a negligible effect for the value of
$\lambda_\text{max}$ we used ($\lambda_\text{max}=4000$).

\section{Comparison with superballs}\label{sec:super_comp}

To compare with superballs we need to relate Batten \emph{et al.}'s aspect ratio $q$~\cite{Batten}
to our $s$. The surface of a superball is described by the equation
\begin{equation}
x^{2q}+y^{2q}+z^{2q}=r^{2q}.
\end{equation}
As mentioned in Ref.~\onlinecite{MM_Patti_cubatic}, from a family of shapes, such as super balls with varying $q$ and $r$, one shape can be selected
that is most similar to the shape of interest, in this case, a spherocube with a certain value of $s$ (the size $l$ sets the length scale:
only $r/l$ is relevant) by minimizing the so-called Hausdorff distance between the two shapes. 

The Hausdorff distance is a distance on shape space that is commonly used in convex geometry.\cite{moszynska2006convex} 
In order to define the Hausdorff distance,
we first define 
\begin{equation}
d'(A,B)=\max_{\mathbf{x}\in A} \min_{\mathbf{y}\in B} \lvert \mathbf{x}-\mathbf{y}\rvert,\label{eqn:Hcoll}
\end{equation}
where $A$ and $B$ are solid (compact) bodies. The Hausdorff distance is then defined by 
\begin{equation}
d(A,B)=\max\{ d'(A,B), d'(B,A)\}.\label{eqn:Hdist}
\end{equation}
In order to calculate the Hausdorff distance we need only consider those sets of points, such that
each set has one point on the surface of $A$ and one on the surface of $B$ and the normals to the respective surfaces
at these points are equal. 

 We only need to consider the distance from the center to the surface in this case in three different symmetry directions.
These distances are listed in Tbl.~\ref{tbl:dists}.
\begin{table}[t!]
\begin{center}
{\addtolength\tabcolsep{1em}
\begin{tabular}{c|cc}
direction  & $d_\text{sb}$ & $d_\text{rc}$\\
\hline
$(1,0,0)$ & $r$ & $l/2$ \\
$(1,1,0)/\sqrt{2}$ & $r \sqrt{2}^{1-1/q}$ & $\sigma/\sqrt{2}+d/2$ \\
$(1,1,1)/\sqrt{3}$ & $r \sqrt{3}^{1-1/q}$ & $\sqrt{3}\sigma/2+d/2$ 
\end{tabular}}
\end{center}
\caption{
The distance between the center of a particle and its surface in the indicated directions  for superballs ($d_{sb}$) and rounded cubes or spherocubes
($d_\text{rc}$).
\label{tbl:dists}}
\end{table}
To calculate the Hausdorff distance, we only need to maximize over the directions $\mathbf{n}$ given in Tbl.~\ref{tbl:dists}:
\begin{equation}
d\big(\text{sb}[r,q],\text{rc}[l,s]\big)=\max_{\mathbf{n}} \lvert d_\text{sb}(r,q)-d_\text{rc}(l,s) \rvert,
\end{equation}
where we use $\text{sb}[r,q]$ and $\text{rc}[l,s]$ to denote a superball and rounded cube, respectively, with a
certain aspect ratio ($s$ or $q$) and linear size ($l$ or $r$).
Minimizing $d\big(\text{sb}[r,q],\text{rc}[l,s]\big)$ for a certain 
value of $s$ with 
respect to $r/l$ and $q$, we obtain the superball that best fits a spherocube with aspect ratio $s$.

\end{document}